\documentclass[single-column,showpacs,groupedaddress]{revtex4}
\usepackage{amsfonts}
\usepackage{amssymb}
\usepackage{amsmath}
\usepackage{graphicx}

\begin{document}

\title{Singularity in the matrix of the coupled Gross-Pitaevskii equations
and the related state-transitions in three-species condensates }
\author{Y.M.Liu$^{1,3}$, Y.Z.He$^2$, and C.G.Bao$^{2*}$}
\affiliation{$^1$Department of Physics, Shaoguan University, Shaoguan, 512005, P. R. China}
\affiliation{$^2$School of Physics, Sun Yat-Sen University, Guangzhou, P. R. China}
\affiliation{$^3$State Key Laboratory of Theoretical Physics, Institute of Theoretical
Physics, Chinese Academy of Sciences, Beijing, 100190, China}

\begin{abstract}
An approach is proposed to solve the coupled Gross-Pitaevskii equations
(CGP) of the 3-species BEC in an analytical way under the Thomas-Fermi
approximation (TFA). It was found that, when the strength of a kind of
interaction increases and crosses over a critical value, a specific type of
state-transition will occur and will cause a jump in the total energy. Due
to the jump, the energy of the lowest symmetric state becomes considerably
higher. This leaves a particular opportunity for the lowest asymmetric state
to replace the symmetric states as the ground state. It was further found
that the critical values are related to the singularity of either the matrix
or a sub-matrix of the CGP. These critical values are not arising from the
TFA but inherent in the CGP, and they can be analytically expressed.
Furthermore, a model (in which two kinds of atoms separated from each other
asymmetrically) has been proposed for the evaluation of the energy of the
lowest asymmetric state. With this model the emergence of the asymmetric
ground state is numerically confirmed under the TFA. The theoretical
formalism of this paper is quite general and can be generalized for BEC with
more than three species.
\end{abstract}

\thanks{Corresponding author: stsbcg@mail.sysu.edu.cn}
\maketitle

\section*{Introduction}

Accompanying the progress in techniques, the research into the 2-species
Bose-Einstein condensates (2-BEC) is gaining an increasing attention in
recent years in both the experimental aspect \cite%
{myat97,ande05,ni08,pilc09,nemi09,wack15,mgr} and theoretical aspect\cite%
{ho96,esry97,pu98,chui99,tripp00,ribo02,chui03,luo07,nott15,
scha15,inde15,kuop15,roy15,luo08,polo15,jsy,mpe,rci,nitta}. Many
distinguished features have been found, say, the existence of various
phases, the critical value in the inter-species interaction and the related
instability, the emergence of asymmetric ground state (g.s.) \cite%
{esry97,chui99,tripp00}, the appearance of vortex, and so on. The study for
BEC with three species (3-BEC) has also started \cite{cal,man,orl,v2}. It is
very interesting to see how the phenomena found in 2-BEC\ would recover in
3-BEC and whether new phenomena would emerge. Note that, for multiband
superconductivity, the interband couplings among a set of different band
condensates are important to the critical behavior of the system. Critical
temperatures are thereby substantially affected (not determined alone by the
Cooper-pair amplitude of a single band)\cite{s1,s2,s3}. Thus, it is
reasonable to expect that the critical phenomena found in 2-BEC (say, a
state-transition has been found to take place when the strength of the
inter-species interaction arrives at a critical value\cite{prep}) might also
be affected and new critical phenomena might emerge. Since the BEC with more
than two species are experimentally achievable, it is meaningful to perform
theoretical research at this stage.

This paper is dedicated to a primary theoretical study on the 3-BEC based on
the coupled Gross-Pitaevskii equations (CGP). Although exact numerical
solutions of the CGP are very valuable, it is not easy to extract the
underlying physics simply via numerical results. In order to gain more
insight into the physics, it is helpful to obtain approximate analytical
solutions. Therefore, the Thomas-Fermi approximation (TFA) has been adopted.
Under the TFA, we provide an approach for obtaining analytical solutions.
Thereby the wave functions and the total energies can be obtained in an
analytical form, and these quantities can relate directly to the parameters
involved. This facilitates greatly related physical analysis. We found that
the singularity of the (sub)matrix-of-equations is crucial to the behavior
of the BEC. Specific state-transitions will be induced when the parameters
vary and cross over a singular point of the matrix. This will be studied in
detail below.

Furthermore, based on the analytical formalism and the singularity of the
equations, effort is made to divide the whole parameter-space into zones,
each supports a specific spatial configuration. This provides a primary
frame for plotting the phase-diagrams in the future. Besides, a model for
calculating the total energies of asymmetric states has also been proposed.
The possibility of the emergence of asymmetric g.s. has been primarily
evaluated.

The theoretical formalism of this paper is quite general and can be
generalized for the K-BEC with K larger than 3.

\section*{Hamiltonian and the coupled Gross-pitaevskii equations}

We assume that the 3-BEC contains $N_{S}$ $S$-atoms with mass $m_{S}$ and
interacting via $V_{S}=c_{S}\sum_{i<i^{\prime }}\delta (\mathbf{r}_{i}-%
\mathbf{r}_{i^{\prime }})$, ($S=A$, $B$ and $C$). The particle numbers are
assumed to be huge (say, $\geq $10000). The inter-species interactions are $%
V_{SS^{\prime }}=c_{SS^{\prime }}\sum_{i=1}^{N_{S}}\sum_{j=1}^{N_{S^{\prime
}}}\delta (\mathbf{r}_{i}-\mathbf{r}_{j})$ with the strength $c_{SS^{\prime
}}$. These atoms are confined by the isotropic harmonic traps $\frac{1}{2}%
m_{S}\omega _{S}^{2}r^{2}$ We introduce a mass $m_{o}$ and a frequency $%
\omega $. Then, $\hbar \omega $ and $\lambda \equiv \sqrt{\hbar
/(m_{o}\omega )}$ are used as units for energy and length. The spin-degrees
of freedom are assumed to be frozen. The total Hamiltonian is
\begin{eqnarray}
H &=&H_{A}+H_{B}+H_{C}+V_{AB}+V_{BC}+V_{CA}  \notag \\
H_{A} &=&\sum_{i=1}^{N_{A}}(-\frac{m_{o}}{2m_{A}}\nabla _{i}^{2}+\frac{1}{2}%
\gamma _{A}r_{i}^{2})+V_{A}  \label{eq1}
\end{eqnarray}%
where $\gamma _{A}=(m_{A}/m_{o})(\omega _{A}/\omega )^{2}$. $H_{B}$ and $%
H_{C}$ are similarly defined.

We assume that no spatial excitations are involved in the g.s.. Thus, each
kind of atoms are fully condensed into a state which is most advantageous
for binding. Accordingly, the total many-body wave function of the g.s. can
be written as
\begin{equation}
\Psi =\prod_{i=1}^{N_{A}}\frac{u_{1}(r_{i})}{\sqrt{4\pi }r_{i}}\
\prod_{j=1}^{N_{B}}\frac{u_{2}(r_{j})}{\sqrt{4\pi }r_{j}}\
\prod_{k=1}^{N_{C}}\frac{u_{3}(r_{k})}{\sqrt{4\pi }r_{k}}  \label{eq2}
\end{equation}%
where $u_{1}$, $u_{2}$ and $u_{3}$ are for the A-, B-, and C-atoms,
respectively.

In the set of the CGP, the one for $u_{1}$ is
\begin{eqnarray}
(-\frac{m_{o}}{2m_{A}}\nabla ^{2}+\frac{1}{2}\gamma _{A}r^{2}+N_{A}c_{A}%
\frac{u_{1}^{2}}{4\pi r^{2}}+N_{B}c_{AB}\frac{u_{2}^{2}}{4\pi r^{2}}\notag \\
+N_{C}c_{CA}\frac{u_{3}^{2}}{4\pi r^{2}}-\varepsilon _{A})u_{1} &=&0 \label{eq3}
\end{eqnarray}%
where $\varepsilon _{A}$ is the chemical potential. Via cyclic permutations
of the three indexes $(A,B,C)$ together with $(u_{1},u_{2},u_{3})$. From eq.(%
\ref{eq3}) we obtain the other two for $u_{2}$ and $u_{3}$. It is emphasized
that the normalization ${\int u_{l}^{2}dr=1}$ ($l$=1, 2 and 3) is required.

\section*{Formal solutions under the Thomas-Fermi approximation}

Since $N_{A}$, $N_{B}$ and $N_{C}$ are considered to be large, TFA has been
adopted. The applicability of this approximation has been evaluated via a
numerical approach given in \cite{yzhe,polo15} and will be discussed later.
Under the TFA, the CGP become

\begin{eqnarray}
(\frac{r^{2}}{2}+\alpha _{11}\frac{u_{1}^{2}}{r^{2}}+\alpha _{12}\frac{%
u_{2}^{2}}{r^{2}}+\alpha _{13}\frac{u_{3}^{2}}{r^{2}}-\varepsilon _{1})u_{1}
&=&0  \notag \\
(\frac{r^{2}}{2}+\alpha _{21}\frac{u_{1}^{2}}{r^{2}}+\alpha _{22}\frac{%
u_{2}^{2}}{r^{2}}+\alpha _{23}\frac{u_{3}^{2}}{r^{2}}-\varepsilon _{2})u_{2}
&=&0  \notag \\
(\frac{r^{2}}{2}+\alpha _{31}\frac{u_{1}^{2}}{r^{2}}+\alpha _{32}\frac{%
u_{2}^{2}}{r^{2}}+\alpha _{33}\frac{u_{3}^{2}}{r^{2}}-\varepsilon _{3})u_{3}
&=&0  \label{eq4}
\end{eqnarray}%
where $\alpha _{11}=N_{A}c_{A}/(4\pi \gamma _{A})$, $\alpha
_{22}=N_{B}c_{B}/(4\pi \gamma _{B})$, $\alpha _{33}=N_{C}c_{C}/(4\pi \gamma
_{C})$, $\alpha _{12}=N_{B}c_{AB}/(4\pi \gamma _{A})$, $\alpha
_{21}=N_{A}c_{AB}/(4\pi \gamma _{B})$, $\alpha _{13}=N_{C}c_{CA}/(4\pi
\gamma _{A})$, $\alpha _{31}=N_{A}c_{CA}/(4\pi \gamma _{C})$, $\alpha
_{23}=N_{C}c_{BC}/(4\pi \gamma _{B})$, $\alpha _{32}=N_{B}c_{BC}/(4\pi
\gamma _{C})$, they are called the weighted strengths (W-strengths) and they
are related as $\alpha _{12}\alpha _{23}\alpha _{31}=\alpha _{21}\alpha
_{32}\alpha _{13}$. $\varepsilon _{1}=\varepsilon _{A}/\gamma _{A}$, $%
\varepsilon _{2}=\varepsilon _{B}/\gamma _{B}$, $\varepsilon
_{3}=\varepsilon _{C}/\gamma _{C}$, they are the weighted energies for a
single particle. Recall that there are originally 15 parameters ($N_{S}$, $%
m_{S}$, $\omega _{S}$, $c_{S}$, $c_{SS^{\prime }}$). Their combined effects
are fully represented by the nine $\alpha _{ll^{\prime }}$ (only eight of
them are independent). Thus, based on the W-strengths, related analysis
could be simpler. In this paper all the interactions are considered as
repulsive. Accordingly, all the W-strengths are positive. Furthermore, it is
safe to assume that all the three $u_{l}/r\geq 0$ (because they do not
contain nodes).

The set of W-strengths forms a matrix $\mathfrak{M}$ (i.e., the
matrix-of-equations) with matrix elements $(\mathfrak{M})_{ll^{\prime
}}=\alpha _{ll^{\prime }}$. The determinant of $\mathfrak{M}$ is denoted by $%
\mathfrak{D}$. The set of equations (\ref{eq4}) has four forms of formal
solutions, each would hold in a specific domain of $r$:

(i) Form III: When all the three wave functions are nonzero in a domain,
they must have the unique form as
\begin{equation}
u_{l}^{2}/r^{2}=X_{l}-Y_{l}r^{2}  \label{eq5}
\end{equation}%
where
\begin{equation}
X_{l}=\mathfrak{D}_{X_{l}}\mathfrak{/D}  \label{eq6}
\end{equation}%
$\mathfrak{D}_{X_{l}}$ is a determinant obtained by changing the $l$ column
of $\mathfrak{D}$ from $(\alpha _{1l},\alpha _{2l},\alpha _{3l})$ to $%
(\varepsilon _{1},\varepsilon _{2},\varepsilon _{3})$.
\begin{equation}
Y_{l}=\mathfrak{D}_{Y_{l}}\mathfrak{/D}  \label{eq7}
\end{equation}%
$\mathfrak{D}_{Y_{l}}$ is also a determinant obtained by changing the $l$
column of $\mathfrak{D}$ to $(1/2,1/2,1/2)$. Once all the parameters are
given, the three $Y_{l}$ are known because they depend only on $\alpha
_{ll^{\prime }}$. However, the three $X_{l}$ have not yet been known because
they depend also on $(\varepsilon _{1},\varepsilon _{2},\varepsilon _{3})$.
When $Y_{l}$\ is positive (negative), $u_{l}/r$\ goes down (up) with $r$.
This point is notable because the main feature of the formal solution
depends on the signs of $\{Y_{l}\}$.

The set $\{X_{l}\}$ and the set $\{\varepsilon _{l}\}$ are related as
\begin{eqnarray}
\varepsilon _{l} &=&\Sigma _{l^{\prime }}\alpha _{ll^{\prime }}X_{l^{\prime
}}  \label{eq7a} \\
X_{l} &=&\Sigma _{l^{\prime }}\overset{\_}{\alpha }_{ll^{\prime
}}\varepsilon _{l^{\prime }}  \label{eq7b}
\end{eqnarray}

where $\overset{\_}{\alpha }_{ll^{\prime }}=\mathfrak{d}_{l^{\prime }l}/%
\mathfrak{D}$, and $\mathfrak{d}_{l^{\prime }l}$ is the algebraic cominor of
$\alpha _{l^{\prime }l}$.

(ii) Form II: When one and only one of the wave functions is zero inside a
domain (say, $u_{n}/r=0$), the other two must have the unique form as
\begin{eqnarray}
u_{l}^{2}/r^{2} &=&X_{l}^{(n)}-Y_{l}^{(n)}r^{2}  \notag \\
u_{m}^{2}/r^{2} &=&X_{m}^{(n)}-Y_{m}^{(n)}r^{2}  \label{eq8}
\end{eqnarray}%
where $l$, $m$, and $n$ are in a cyclic permutation of 1-2-3 (the same in
the follows),
\begin{eqnarray}
X_{l}^{(n)} &=&(\alpha _{mm}\varepsilon _{l}-\alpha _{lm}\varepsilon _{m})/%
\mathfrak{d}_{nn}  \notag \\
Y_{l}^{(n)} &=&\frac{1}{2}(\alpha _{mm}-\alpha _{lm})/\mathfrak{d}_{nn}
\notag \\
X_{m}^{(n)} &=&(\alpha _{ll}\varepsilon _{m}-\alpha _{ml}\varepsilon _{l})/%
\mathfrak{d}_{nn}  \notag \\
Y_{m}^{(n)} &=&\frac{1}{2}(\alpha _{ll}-\alpha _{ml})/\mathfrak{d}_{nn}
\label{eq9}
\end{eqnarray}%
Once the parameters are given, the six $Y_{n^{\prime }}^{(n)}$ ($n^{\prime
}\neq n$) are known, while the six $X_{n^{\prime }}^{(n)}$ have not yet.
When $Y_{n^{\prime }}^{(n)}$\ is positive (negative), $u_{n^{\prime }}/r$\
goes down (up) with $r$. When the Form II has $u_{n}/r=0$, a more precise
notation Form II$_{n}$ is adopted for the detail.

(iii) Form I: When one and only one of the wave functions is nonzero in a
domain (say, $u_{l}/r\neq 0$), it must have the unique form as
\begin{equation}
u_{l}^{2}/r^{2}=\frac{1}{\alpha _{ll}}(\varepsilon _{l}-r^{2}/2)
\label{eq10}
\end{equation}%
Obviously, $u_{l}/r$ in this form must descend with $r$. For the case $%
u_{l}/r\neq 0$, the more precise notation Form I$_{l}$ is adopted.

(iv) Form 0: In this form\ all the three wave functions are zero.

If a wave function (say, $u_{l}/r$) is nonzero in a domain but becomes zero
when $r=r_{o}$, then a downward form-transition (say, from Form III to II)
will occur at $r_{o}$. Whereas if $u_{l}/r$\ is zero in a domain but becomes
nonzero when $r=r_{o}$, then a upward form-transition (say, from Form II to
III) will occur at $r_{o}$. $r_{o}$ is named a form-transition-point, and it
appears as the boundary separating the two connected domains. In this way
the formal solutions serve as the building blocks, and they will link up
continuously to form an entire solution. They must be continuous at the
form-transition-points because the wave functions satisfy exactly the same
set of nonlinear equations at those points. However, their derivatives are
in general not continuous at the boundaries.

\section*{An approach for obtaining analytical solutions of the CGP}

In this section we consider the case that all the parameters are given and
the values of the three $\{{\varepsilon _{l}\}}$ have been presumed. In this
case all the formal solutions are known. We will propose an approach to link
up the formal solutions to form a chain as a candidate of an entire
solution. To this aim we first introduce a number of features related to the
linking.

(i) For Form I to III, at least one of the wave function is descending with $%
r$.

The proof of this feature is referred to \cite{liu}, where it is proved that
at least one of $Y_{l}$ (or $Y_{l}^{(n)}$ for a given $n$) is positive.

This feature implies that, when $r$\ increases, the occurrence of a downward
form-transition is inevitable, unless a upward form-transition takes place
prior to the downward transition. In any cases a formal solution must
transform to another form somewhere (except Form 0).

(ii) For a formal solution existing in a domain, the right boundary of the
domain and the successor (the successive formal solution) in the next domain
have been prescribed when the three $\{{\varepsilon _{l}\}}$ have been
presumed.

To prove this feature, as an example, we assume that $u_{1}/r$ and $u_{3}/r$
are nonzero in a domain while $u_{2}/r$ is zero. This assumption implies
that we have assumed $X_{1}^{(2)}-Y_{1}^{(2)}r^{2}\geq 0$ and $%
X_{3}^{(2)}-Y_{3}^{(2)}r^{2}\geq 0$ when $r$\ is given inside the domain
(refer to eq.(\ref{eq8})). We define $r_{1}^{2}=X_{1}^{(2)}/Y_{1}^{(2)}$ or $%
\infty $ (if $Y_{1}^{(2)}>0$ or $\leq 0)$. Similarly, we define $%
r_{3}^{2}=X_{3}^{(2)}/Y_{3}^{(2)}$ or $\infty $ (if $Y_{3}^{(2)}>0$ or $\leq
0$), and $r_{2}^{2}=X_{2}/Y_{2}$ or $\infty $ (if both $X_{2}$ and $Y_{2}$
are negative or otherwise). Then, the smallest one among $r_{1}$, $r_{2}$,
and $r_{3}$ is just the right boundary of the domain. Say, if $r_{1}$\ is
the smallest, then $u_{1}/r\rightarrow 0$ when $r\rightarrow r_{1}$, and the
successor will have the Form I$_{3}$. If $r_{2}$\ is the smallest, then $%
u_{2}/r$ will emerge at $r_{2}$, and the successor will have the Form III,
and so on. Since\ $r_{1}$, $r_{2}$, and $r_{3}$ are prescribed, the right
boundary and the successor are prescribed

(iii) Once the formal solution in the first domain (starting from $r=0$) is
prescribed, the formal solutions will link up one-by-one to form a chain in
a unique way. There are seven types of formal solutions (say, in Form II$%
_{2} $ , or in Form I$_{3}$, and so on). Each type can appear in a chain at
most once.

Obviously, since the successor in each step of linking is uniquely
prescribed, the whole chain is prescribed. Since the right boundary of a
type is prescribed, the type can not appear twice.

(iv) Once a formal solution in a chain is in Form 0, the chain will end.

This is because no wave functions can emerge from an empty domain.
Otherwise, if $u_{1}/r$ emerges alone, it must have the form as eq.(\ref%
{eq10}). This form prohibits the uprising of $u_{1}/r$. If $u_{1}/r$ and $%
u_{2}/r$\ emerge at the same place, $Y_{1}^{(3)}$ and $Y_{2}^{(3)}$ must
both be negative. This violates the feature (i).\quad If\ all the $%
\{u_{l}/r\}$ emerge at the same place, all the three \{$Y_{l}$\} must be
negative. This violates also the feature (i).

Based on the above features, we propose an approach as follows: First, we
design a chain for a type of entire solutions denoted as, for an example, II$%
_{2}$-III-II$_{1}$-I$_{3}$ (it implies that the first domain has a Form II$%
_{2}$, the next domain has a Form III, the third domain has a Form II$_{1}$,
while the last domain has a Form I$_{3}$). The prescription on the linking
appears as a number of requirements (inequalities) imposing on the
W-strengths and the presumed $\{\varepsilon _{l}\}$. When all the $\{\alpha
_{ll^{\prime }}\}$ and the $\{\varepsilon _{l}\}$ are given inside a
specific scope, all the requirements can be met and the designed chain as a
candidate can be achieved. At this stage the normalization has not yet been
considered. When the three equations $\int u_{l}^{2}dr=1$ are further
introduced, not only the scope but the values of the set $\{\varepsilon
_{l}\}$ can be fixed. Then, the candidate will be a realistic entire
solution of the CGP. In general, the three equations can uniquely determine
the three unknowns $\{\varepsilon _{l}\}$, unless the design itself is not
reasonable. Thus, when the parameters are given in a reasonable scope, we
can uniquely find out a realistic entire solution, which is a chain of
formal solutions with a specific linking. A detailed practice of this
approach for miscible states is given in \cite{liu}.

\section*{State-transition and the singularity of the matrix}

Based on the above approach, numerical calculations for two types of
examples are performed. Related wave functions are plotted.

(1) State-transition occurring at the singular point of the
matrix-of-equations

\begin{figure}[tbp]
\centering \resizebox{0.95\columnwidth}{!}{\includegraphics{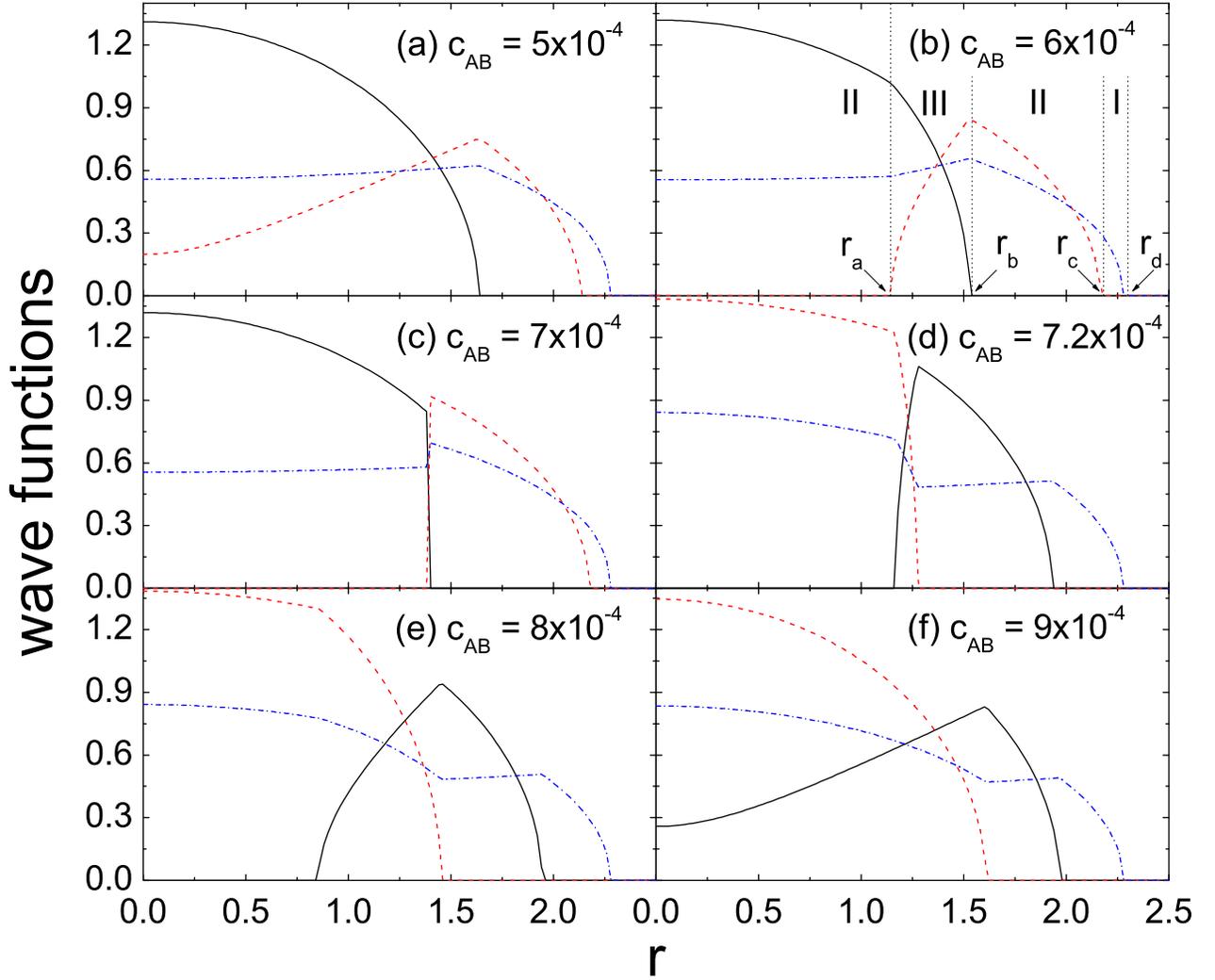} }
\caption{(color online) Wave functions $u_1/r$ (in solid line), $u_2/r$ (in
dash line), and $u_3/r$ (in dash-dot line) against $r$. $\hbar \protect%
\omega $ and $\protect\lambda\equiv\protect\sqrt{\hbar/(m_o\protect\omega)}$
are used as units for energy and length. $c_{AB}$ is given at six values
marked in the panels. Other parameters are fixed and are chosen quite
arbitrary but having $p_{BC}$ and $p_{CA}$ both being negative. They are $%
N_A=30000$, $N_B=11000$, $N_C=29000$, $c_A=4\times 10^{-4}$ (in $\hbar%
\protect\omega\protect\lambda^3$, the same for other strengths), $%
c_B=1.4\times 10^{-3}$, $c_C=1.2\times 10^{-3}$, $c_{BC}=3.8\times 10^{-4}$,
$c_{CA}=4.2\times 10^{-4} $, $\protect\gamma_A=\protect\gamma_B=\protect%
\gamma_C=1$. In 1b, the Forms of the solutions are marked in the associated
domains separated by the vertical dotted lines.}
\label{Fig:1}
\end{figure}

Fig.1 (b) to (e) is for II$_{2}$-III-II$_{1}$-I$_{3}$, while (a) and (f) are
for III-II$_{1}$-I$_{3}$. From (a) to (f) $c_{AB}$ is increasing while the
other parameters remain unchanged. Thus these patterns demonstrate the effect
of $c_{AB}$.

For 1b as an example, the second domain is in Form III. For this form both $%
\{X_{l}\}$ and $\{Y_{l}\}$ are proportional to $1/\mathfrak{D}$. Therefore,
when $\mathfrak{D}\rightarrow 0$, the wave functions will become extremely
steep and the second domain will become extremely narrow as shown in (c). It
turns out that, when $c_{AB}=7.02\times 10^{-4}\hbar \omega \lambda
^{3}\equiv c_{AB}^{crit(3)}$, the matrix-of-equations becomes singular and
accordingly $\mathfrak{D}=0$. When $c_{AB}$\ is close to and crosses over
this critical value (from (c) to (d)), $\{Y_{l}\}$ will suddenly change
their signs. It implies a down-falling wave function suddenly becomes
up-rising, and accordingly the whole pattern is changed greatly. This is
definitely associated with a state-transition in which all the A-atoms
suddenly jump from a core to a shell, while all the B-atoms jump in a
reverse way as clearly shown in (c) and (d). Accompanying the great change,
a remarkable increase in the total energy is expected (this expectation is
confirmed below).

From the equality $\mathfrak{D}=0$, it is straight forward to obtain
\begin{equation}
c_{AB}^{crit(3)}=\frac{1}{c_{C}}(c_{BC}c_{CA}\pm \sqrt{p_{BC}p_{CA}})
\label{eqcab}
\end{equation}

where
\begin{equation}
p_{BC}\equiv c_{BC}^{2}-c_{B}c_{C}  \label{eqpbc}
\end{equation}

Similarly, $p_{CA}$ and $p_{AB}$ can be defined by permuting the indexes.

Note that:

(i) When $p_{BC}p_{CA}<0$, $c_{AB}^{crit(3)}$ does not exist (i.e., the
matrix will not become singular). Therefore, even a Form III is contained in
a chain, the variation of $c_{AB}$ does not assure the occurrence of the
state-transition. Only if the other five strengths are so chosen that $%
p_{BC}p_{CA}\geq 0$, the critical point could exist and the transition could
occur.

(ii) $c_{AB}^{crit(3)}$ deviates remarkably from the well known critical
value $c_{AB}^{crit(2)}=\sqrt{c_{A}c_{B}}$ for 2-BEC. Thus, the
state-transition is remarkably affected by the influence of the third kind
of atoms. However, if both $c_{BC}\rightarrow 0$ and $c_{CA}\rightarrow 0$
(i.e., the influence is removed), one can prove from eq.(\ref{eqcab}) that $%
c_{AB}^{crit(3)}\rightarrow c_{AB}^{crit(2)}$.

(iii) $c_{AB}^{crit(3)}$ depends on the other five strengths but not on the
particle numbers, trap frequencies, and masses. This feature is the same as
what has found in 2-BEC. Thus, in an experiment, the variation of the
parameters other than the strengths will not change the critical values.

(iv) The state-transitions caused by the variation of other strengths can be
similarly deduced. For an example, for the intra-species interaction of the
A-atoms, when $c_{A}$\ increases and arrives at a critical value
\begin{equation}
c_{A}^{crit(3)}=\frac{1}{P_{BC}}%
(c_{CA}(c_{AB}c_{BC}-c_{B}c_{CA})-c_{AB}(c_{AB}c_{C}-c_{BC}c_{CA}))
\label{eqca}
\end{equation}%
the matrix will become singular and the transition will occur.

In summary, for an entire solution contains a Form III, when the variation
of the strengths leads to a cross-over of the singular point of the matrix,
a state-transition will occur. Since the singularity of the matrix is
inherent in the CGP but not a product of the TFA, thus the occurrence of the
state-transitions at the critical values holds beyond the TFA. In fact, in
the earliest study of the 2-BEC, the instability in the neighborhood of the
critical value $c_{AB}^{crit(2)}=\sqrt{c_{A}c_{B}}$ (the singular point of
the two-rank matrix) has been pointed out \cite{ho96}.

(2) State-transition occurring at the singular point of a
sub-matrix-of-equations

When an entire solution contains a Form II, another type of state-transition
might occur. In Fig.2, the entire solution is III-II$_{3}$-I$_{2}$ in (a) to
(c), and is III-II$_{3}$-I$_{1}$ in (d) to (f),\ where the second domain has
the Form II$_{3}$ (namely, only the A- and B-atoms are contained in this
domain). For the Form II$_{3}$ the critical value of $c_{AB}$\ is\ $%
c_{AB}^{crit(2)}=\sqrt{c_{A}c_{B}}=8.944\times 10^{-4}\hbar \omega \lambda
^{3}$, which is the singular point of the sub-matrix of the equations for $%
u_{1}/r$ and $u_{2}/r$ only. When $c_{AB}$\ is\ close to this value ((c) and
(d)) the second domain becomes very narrow, and the two wave functions\
become very steep. During the cross-over, $Y_{1}^{(3)}$ and $Y_{2}^{(3)}$
change their signs and a transition occurs as shown in (c) and (d).
Nonetheless, different from the one found in Fig.1, only a part of the A-
and B-atoms are actively taking part in this transition, namely, a part of
A-atoms rush out from the core and form a shell, while a part of outward
B-atoms rush from the shell into the core. Thus, the corresponding change in
spatial configuration is relatively milder. The change appears essentially
in the second and the third domains where the C-atoms are absent.
Accordingly, the critical value is not at all affected by the C-atoms and is
identical to the value of 2-BEC. Incidentally, although the Form III is
contained in Fig.2, $c_{AB}^{crit(3)}$ does not exist in this case due to $%
p_{BC}p_{CA}<0$.

\section*{Total energy of symmetric states and the great jump}

When the total energy of the lowest symmetric state $E_{tot}$\ is higher
than the total energy of the lowest asymmetric state $E_{tot}^{asym}$, the
g.s. will be asymmetric. Thus, $E_{tot}>E_{tot}^{asym}$ is the discriminant
to judge whether the g.s. is asymmetric.

When the wave functions are known we can obtain the total energy as (the
kinetic energy has been omitted)
\begin{equation}
E_{tot}=\sum_{i}(P_{i}+E_{i})+\Sigma _{i<i^{\prime }}E_{ii^{\prime }}
\label{eq20}
\end{equation}%
where $i=1$, 2 and 3. They are associated with $A-$, $B-$, and $C-$atoms,
respectively. $P_{1}=\frac{N_{A}\gamma _{A}}{2}\int u_{1}^{2}r^{2}dr$, $%
E_{1}=\frac{N_{A}^{2}c_{A}}{8\pi }\int (u_{1}/r)^{4}r^{2}dr$, $E_{12}=\frac{%
N_{A}N_{B}c_{AB}}{4\pi }\int (u_{1}/r)^{2}(u_{2}/r)^{2}r^{2}dr$, and so on.
Let $N=N_{A}+N_{B}+N_{C}$. Examples of $E_{tot}/N$ versus $c_{AB}$\ are
plotted via the solid lines shown in Fig.3. The other parameters in (a) and
(b) are the same as in Fig.1 and Fig.2, respectively. A distinguished
feature is the appearance of the great jump at $c_{AB}^{crit(3)}$ (a) and $%
c_{AB}^{crit(2)}$ (b). Note that, in Fig.3(a), the crossing over $%
c_{AB}^{crit(2)}$\ does not cause an effect because the associated
transition could occur only if the chain contains the building block II$_{3}$%
, this building block is absent in Fig.1. While in Fig.3(b) $c_{AB}^{crit(3)}
$ does not exist because $p_{BC}p_{CA}<0$ as mentioned.

Fig.3 confirms that the state-transition has caused a great change in $%
E_{tot}$. For the transition shown in Fig.1(c) and (d), when the B-atoms
rush in, $E_{2}$ will increase (because a more compact distribution leads to
the increase of the factor $\int (u_{2}/r)^{4}r^{2}dr$)\ while $P_{2}$\ will
remarkably decrease. The decrease over takes the increase. We found that,
for each B-atom, $(E_{2}+P_{2})/N_{B}$ decreases from 1.80 to 1.41. On the
other hand, for each A-atom, $(E_{1}+P_{1})/N_{A}$ increases from 1.08 to
1.50. Since $N_{A}>>N_{B}$ in this example, totally, $E_{tot}$ increases
remarkably. This examples demonstrates that, although the critical value for
the transition depends only on the strengths, the magnitude of the energy
gap depends also on other parameters. The magnitude can be very large or
quite small (say, in the above examples, the magnitude can be tuned by
varying $N_{A}$ and/or $N_{B}$). Since all the A- and B- atoms are involved
in the transition, the excitation is collective in nature.

\section*{Asymmetric states and their total energy}

We know from the study of the 2-BEC \cite{esry97,chui99,tripp00} that, when $%
V_{AB}$ is sufficiently strong, the A- and B-atoms might give up the
symmetry of the trap for lowering the g.s. energy. Therefore, we propose a
model where only the distributions of the A- and B-atoms are asymmetric,
while the C-atoms are symmetric. Let $O$ denotes the center of the trap. Let
a sphere with radius $R_{AB}$ centered at $O$ be divided into two parts by a
plane perpendicular to the Z-axis. The plane intersects the Z-axis at $%
z=z_{0}$ ($-R_{AB}<z_{0}<R_{AB}$). Let the A-atoms be distributed in the
lower part of the sphere, and the B-atoms in the upper part. Let the C-atoms
be symmetrically distributed in another sphere with radius $R_{C}$ and
centered also at $O$. Then, we assume
\begin{equation}
u_{1}/r=d_{1}\sqrt{1-(r/R_{AB})^{2}}  \label{u1}
\end{equation}%
if $r\leq R_{AB}$ and $z\leq z_{0}$. Otherwise, it is zero. Where $d_{1}=(%
\frac{1}{15}R_{AB}^{3}+\frac{1}{8}R_{AB}^{2}z_{0}-\frac{1}{12}z_{0}^{3}+%
\frac{1}{40}R_{AB}^{-2}z_{0}^{5})^{-1/2}$ is for the normalization.
\begin{equation}
u_{2}/r=d_{2}\sqrt{1-(r/R_{AB})^{2}}  \label{u2}
\end{equation}%
if $r\leq R_{AB}$ and $z>z_{0}$. Otherwise, it is zero. Where $d_{2}=(\frac{1%
}{15}R_{AB}^{3}-\frac{1}{8}R_{AB}^{2}z_{0}+\frac{1}{12}z_{0}^{3}-\frac{1}{40}%
R_{AB}^{-2}z_{0}^{5})^{-1/2}$ and
\begin{equation}
u_{3}/r=d_{3}\sqrt{1-(r/R_{C})^{2}}  \label{u3}
\end{equation}

if $r\leq R_{C}$. Otherwise, it is zero. Where $d_{3}=(\frac{2}{15}%
R_{C}^{3})^{-1/2}$ When the values of $R_{AB}$, $R_{C}$, and $z_{0}$ are
assumed, from eqs.(\ref{u1},\ref{u2},\ref{u3}) and eq.(\ref{eq20}), the
total energy for the asymmetric state (with the kinetic energies neglected),
denoted as $E_{tot}^{asym}$, can be obtained. The parameters $R_{AB}$, $R_{C}
$, and $z_{0}$ are considered as variable. Eventually, they fixed at the
values that lead to the minimum of $E_{tot}^{asym}$. The $E_{tot}^{asym}$
obtained via such a variational procedure is in general higher than its
actual value. Thus, in any cases, if we found $E_{tot}^{asym}<E_{tot}$, the
asymmetric state will inevitably replace the symmetric g.s..

The comparison of the two energies are shown in Fig.3, where (a) and (b) are
associated with Fig.1 and Fig.2, respectively.

\begin{figure}[tbp]
\centering \resizebox{0.95\columnwidth}{!}{\includegraphics{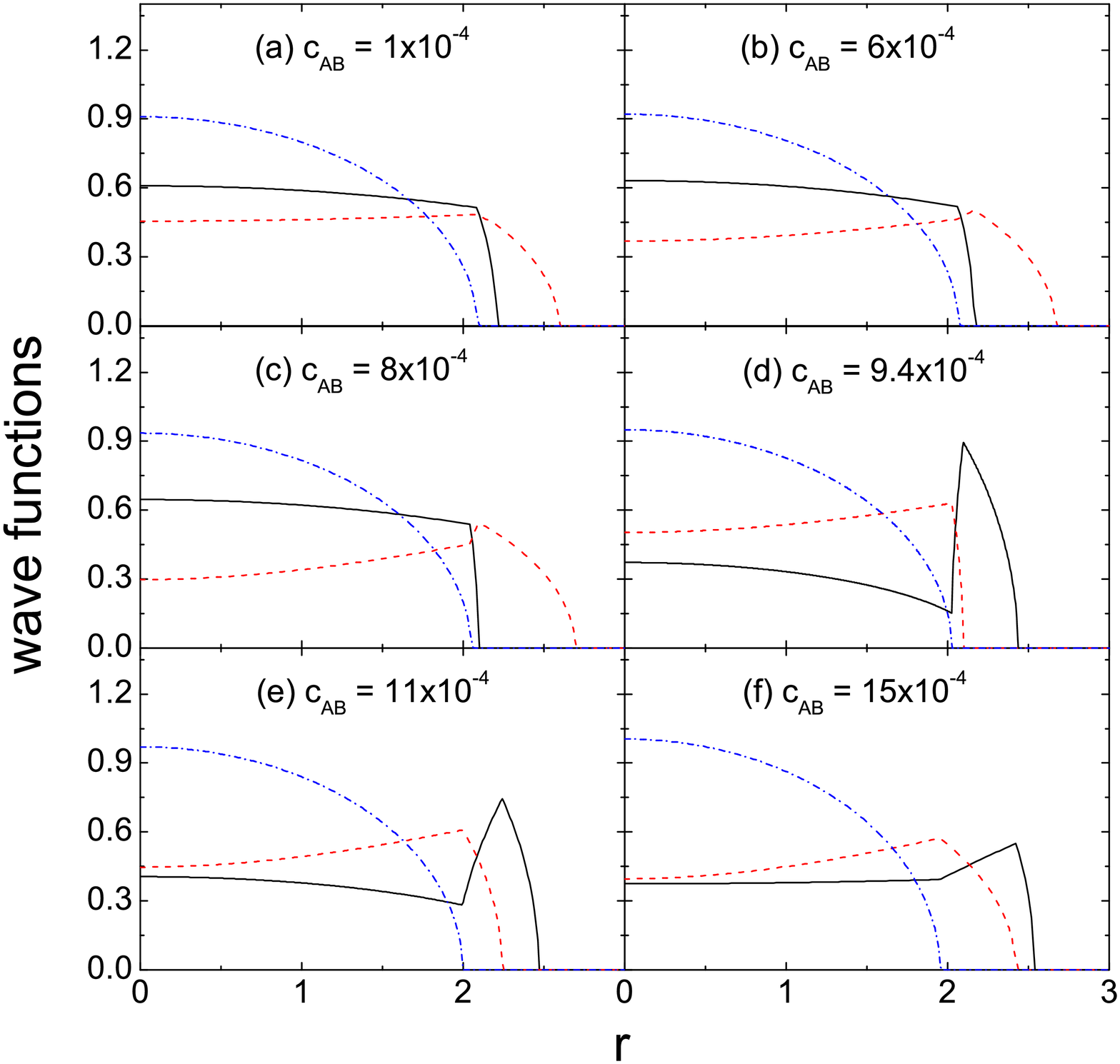} }
\caption{(color online) The same as in Fig.1 but the parameters are so given
that $p_{BC}$ and $p_{CA}$ are in opposite signs. The details of parameters
are $N_A=N_B=N_C=30000$, $c_A=4\times 10^{-4}$ (in $\hbar\protect\omega%
\protect\lambda^{3}$, the same in the follows), $c_B=2\times 10^{-3}$, $%
c_C=1\times 10^{-3}$, $c_{BC}=11.5\times 10^{-4}$, $c_{CA}=10.5\times
10^{-4} $, $\protect\gamma_A=\protect\gamma_B=\protect\gamma_C=1$.}
\label{Fig.2}
\end{figure}

\begin{figure}[tbp]
\centering \resizebox{0.95\columnwidth}{!}{\includegraphics{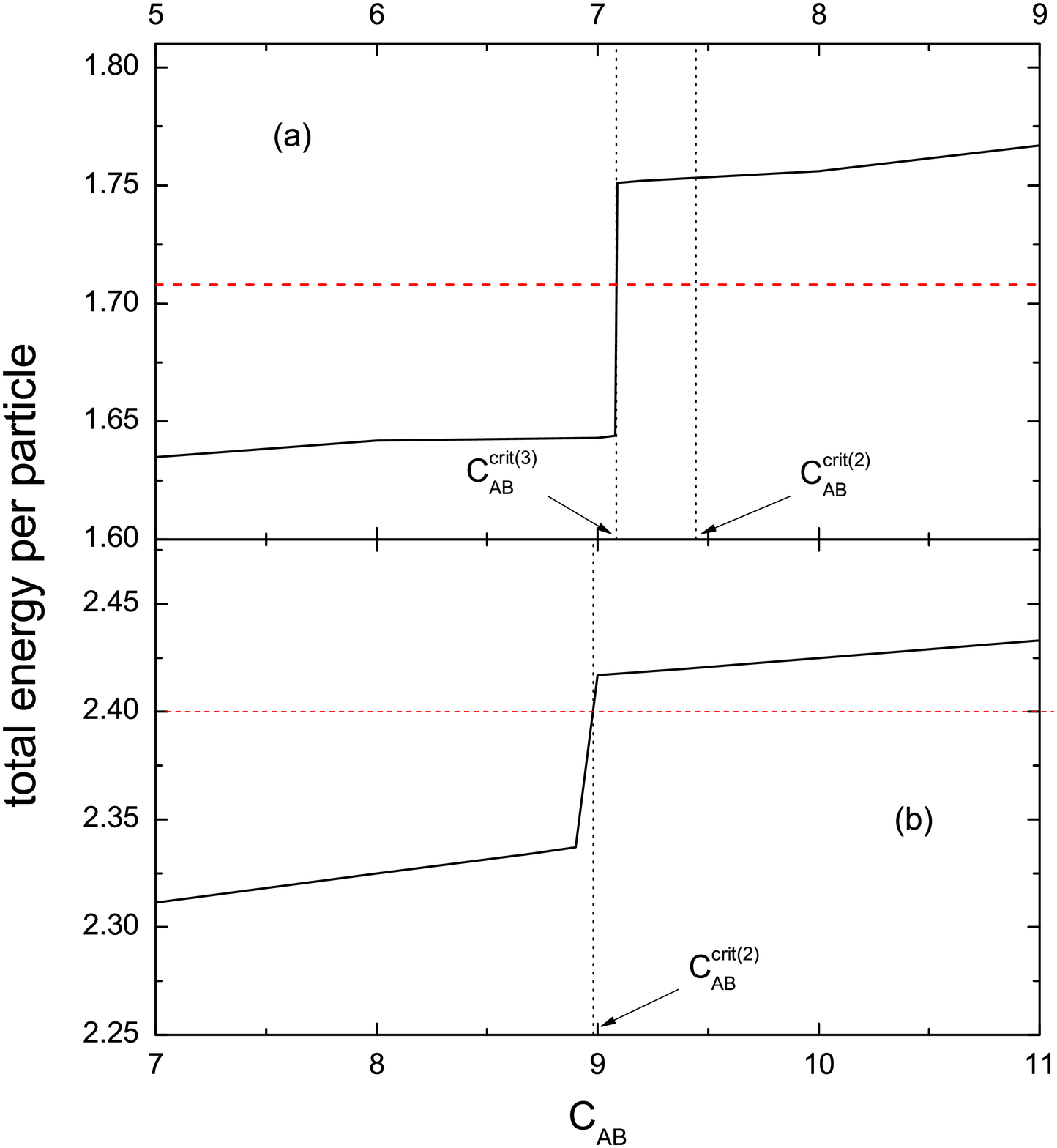} }
\caption{(color online), $E_{tot}/N$ (in solid line for the lowest symmetric
state) and $E_{tot}^{asym}/N$ (in horizontal dash line for the lowest
asymmetric state) versus $c_{AB}$. Other parameters in (a) and (b) are the
same as in Fig.1 and Fig.2, respectively. The unit $\hbar \protect\omega $
is used for energy, and $10^{-4}\hbar \protect\omega \protect\lambda ^{3}$
for $c_{AB}$. Note that the ranges of $c_{AB}$ in (a) and (b) are different.}
\label{Fig.3}
\end{figure}

Fig.3a demonstrates clearly that $E_{tot}^{asym}$ is remarkably lower than $%
E_{tot}$ when $c_{AB}>c_{AB}^{crit(3)}$. Thus the jump provides a good
opportunity for the lowest asymmetric state to replace the symmetric g.s..
Whereas when $c_{AB}<c_{AB}^{crit(3)}$, although $E_{tot}^{asym}$ is
remarkably higher than $E_{tot}$ as shown in the figure and $E_{tot}$ will
decrease further with the decrease of $c_{AB}$, we can only say that the
g.s. is very probable to be symmetric. This is a point to be further studied.

\section*{Division of the parameter-space}

If the whole parameter-space $\Sigma $, in which a point is associated with
a set of parameters, can be divided into zones each supports a specific
configuration, various phase-diagrams could be plotted. Thereby the
essential features of the system and the effects of the parameters can be
visualized. Due to having so many parameters, the phase-diagrams of a 3-BEC\
would be very complicated. At this moment we are not able to plot them. The
following is a primary attempt along this line.

There are four well defined and important surfaces in $\Sigma $. They are
expressed via the equations $\mathfrak{D}=0$ and $\mathfrak{d}_{ii}=0$ ($i=$%
1 to 3). In other words, each surface is an aggregation of a kind of
singular points. We have proved under the TFA that a crossing over these
surfaces may cause a state-transition and accordingly an increase of $%
E_{tot} $. When the TFA is removed, in a domain of $r$\ in which all the $%
\{u_{i}/r\} $\ are nonzero, the exact CGP can be written as
\begin{equation}
\left(
\begin{array}{c}
u_{1}^{2}/r^{2} \\
u_{2}^{2}/r^{2} \\
u_{3}^{2}/r^{2}%
\end{array}%
\right) =\frac{1}{\mathfrak{D}}\left(
\begin{array}{c}
\mathfrak{d}_{11},\mathfrak{d}_{21},\mathfrak{d}_{31} \\
\mathfrak{d}_{12},\mathfrak{d}_{22},\mathfrak{d}_{32} \\
\mathfrak{d}_{13},\mathfrak{d}_{23},\mathfrak{d}_{33}%
\end{array}%
\right) \left(
\begin{array}{c}
\varepsilon _{1}-\frac{r^{2}}{2}+\frac{1}{2}(\frac{m_{o}\omega }{m_{A}\omega
_{A}})^{2}\frac{u_{1}^{"}}{u_{1}} \\
\varepsilon _{2}-\frac{r^{2}}{2}+\frac{1}{2}(\frac{m_{o}\omega }{m_{B}\omega
_{B}})^{2}\frac{u_{2}^{"}}{u_{2}} \\
\varepsilon _{3}-\frac{r^{2}}{2}+\frac{1}{2}(\frac{m_{o}\omega }{m_{C}\omega
_{C}})^{2}\frac{u_{3}^{"}}{u_{3}}%
\end{array}%
\right)  \label{eqm}
\end{equation}

where $u_{i}"$\ is the second-order derivative of $u_{i}$\ against $r$. The
appearance of the common factor $1/\mathfrak{D}$\ at the right side implies
that the left-side (namely, the wave functions) is extremely sensitive
against the parameters when they are given in the neighborhood of the
surface $\mathfrak{D}=0$. This is an important feature of the CGP. When a
point in $\Sigma $ crosses over $\mathfrak{D}=0$, the factor $1/\mathfrak{D}$%
\ changes from $\mp \infty $ to $\pm \infty $. Therefore, the entire
solutions (if it contains a Form III) will undergo a dramatic change, and
the occurrence of the state-transition (found before under the TFA, refer to
Fig.1) is inevitable Thus, this kind of transition is inherent in the CGP.
For the kind of entire solutions containing a Form III, once the variation
of the parameters leads to a crossing over the surface $\mathfrak{D}=0$, the
transition (denoted as trans-III) happens definitely.

Similarly, in a domain of $r$\ in which $u_{n}/r=0$, the exact CGP can be
written in a form in which both $(u_{l}/r)^{2}$ and $(u_{m}/r)^{2}$ are
proportional to a common factor $1/\mathfrak{d}_{nn}$. Thus, for the type of
entire solutions containing a Form II$_{n}$,\ the crossing over the surface $%
\mathfrak{d}_{nn}=0$\ will also lead to a great change in $u_{l}/r$ and $%
u_{m}/r$, and accordingly another kind of state-transition (denoted as
trans-II$_{n}$) occurs as shown in Fig.2.

Let us define a subspace $\Sigma _{III}$\ as follows. When a set of
parameters leads to an entire solution containing a Form III, then the
associated point belongs to $\Sigma _{III}$, otherwise belongs to its
complement. Let the part of the surface $\mathfrak{D}=0$ located inside $%
\Sigma _{III}$\ be denoted as $\sigma _{III}$. Then, $\sigma _{III}$ appears
as a boundary, the crossing over this boundary leads to the trans-III.
Similarly, let $\Sigma _{II_{3}}$ denotes the subspace containing the points
each leads to an entire solution containing a Form II$_{3}$. Let the part of
the surface $\mathfrak{d}_{nn}=0$\ located inside the subspace $\Sigma
_{II_{3}}$ be denoted as $\sigma _{II_{3}}$. Then, the crossing over $\sigma
_{II_{3}}$ leads to the transition trans-II$_{3}$. We can further define $%
\sigma _{II_{1}}$ and $\sigma _{II_{2}}$ in a similar way. These four
surfaces ($\sigma _{III}$ and the three $\sigma _{II_{i}}$) together form
the boundaries and provide a primitive division of $\Sigma $. At the two
sides of each boundary, the entire solutions are greatly different.

Nonetheless, these boundaries are not the actual boundaries for the
phase-diagrams of the g.s.. The latter can be made certain only if exact
calculations on both the lowest symmetric and asymmetric states have been
performed. However, since the crossing over the above boundaries leads to an
increase of $E_{tot}$ and the increase may be large (as shown in Fig.3).
Thus the increase provides an excellent opportunity for the lowest
asymmetric state to replace the lowest symmetric state and become the g.s..
Therefore, we believe that the exact boundaries for the phase diagrams would
partially overlap the boundaries from singularity.

\section*{Final remarks}

(1) A general approach is proposed to solve the CGP for 3-BEC in an
analytical way. TFA has been adopted. The essence of this approach is to
find out the building blocks, i.e., the formal solutions, and the rules for
their linking. The entire solutions of the CGP appear as a chain of them.
This approach is applicable for obtaining solutions with their chains in
various types, and can be generalized to K-BEC with K larger than three. For
examples, in a domain where all the K $\{u_{l}/r\}$ are nonzero, the formal
solution has exactly the same expressions as shown in eqs.(\ref{eq5},\ref%
{eq6},\ref{eq7}) except that the related matrixes are K-rank.

(2) The main result of this paper is the finding of the state-transitions
caused by the singularity of the (sub)matrix-of-equation and the associated
increase of $E_{tot}$ during the transition. The singularity is not a
by-product of the TFA, but an important feature inherent in the CGP. Note
that the critical behavior of the multiband superconductors was found to be
substantially affected by the interband coupling\cite{s1,s2,s3}. Similarly, the
critical point for the state-transition found in this paper differs
remarkably from the one of the 2-BEC (refer to Fig.1) due to the
inter-species coupling. Note that the 3-BEC contains three subsystems, each
contains two species. Similar to the hidden criticality found also in
multiband superconductivity\cite{s2}, the critical points of the three
subsystems appear as the hidden critical points of the 3-BEC. Under specific
conditions state-transitions will also occur at these hidden points (refer
to Fig.2).

(3) A model for asymmetric states has been proposed. Via numerical
calculations on some examples, it is demonstrated that the lowest asymmetric
state replaces the symmetric states and become the g.s. when the strength of
an inter-species interaction arrives at and exceeds its critical value.

(4) The whole parameter-space is primitively divided into zones separated by
four surfaces as boundaries, each is an aggregation of a kind of singular
points. The spatial configurations at the two sides of a boundary are
greatly different due to the state-transition occurring during the crossing
over the boundaries. The transition is accompanied with an energy increase,
the amount of increase might be very large. Thus the state-transition
provides an excellent opportunity for the emergence of the asymmetric g.s..
Therefore, it is expected that the exact boundaries designating the zones of
asymmetric g.s. would overlap partially with the boundaries arising from the
singularity. This remains to be confirmed.

\bigskip

\begin{acknowledgments}
Supported by the National Natural Science Foundation of China under Grants
No.11372122, 11274393, 11574404, and 11275279; the Open Project Program of
State Key Laboratory of Theoretical Physics, Institute of Theoretical
Physics, Chinese Academy of Sciences, China(No.Y4KF201CJ1); and the National
Basic Research Program of China (2013CB933601).
\end{acknowledgments}

\newpage

Additional Information

Author contribution statement

Y.M. Liu is responsible to the theoretical derivation.

Y.Z. He is responsible to the numerical calculation.

C.G. Bao provides the idea, write the paper, and responsible to the whole
paper.

\bigskip Competing Financial Interests

The authors declare no competing financial interests.

\end{document}